\documentclass[useAMS,usenatbib]{mn2e}
\usepackage{times}
\usepackage{psfig}
\title[Host galaxies and black hole masses of RLAGN]
{Host galaxies and black hole masses of low and high luminosity radio loud active nuclei.}
\author[R. Falomo et al.]{ Renato Falomo$^1$,
Nicoletta Carangelo$^{2,3}$
        and
       Aldo Treves $^2$\\ 
   $^1$Osservatorio Astronomico di Padova, 
   Vicolo dell'Osservatorio 5, 35122 Padova, Italy \\
 $^2$Universita' dell'Insubria, via Valleggio 11, 22100 Como, Italy\\ 
$^3$Universita' di Milano-Bicocca, Piazza della Scienza 3, 20126 Milano, Italy \\}
\begin{document}

\date{Accepted .... Received...  }

\pagerange{\pageref{firstpage}--\pageref{lastpage}} \pubyear{200?}

\maketitle

\label{firstpage}

\begin{abstract}
We investigate the host galaxy luminosities of  BL Lac Objects (BLLs) and Radio Loud
Quasars (RLQs) at z$<$0.5 imaged with the Hubble Space Telescope (HST). From a
homogeneous treatment of the data we construct the host galaxy luminosity functions
(HGLFs) and find that RLQ hosts are $\sim$ 0.5 mag brighter than those of BLL: $<M_{R}>
_{RLQ}$ = --24.0, $<M_{R}>_{BLL}$= --23.5. For both classes the HGLFs exhibit a
remarkably different distribution with respect to that of normal (inactive) ellipticals,
with clear preference for more luminous  galaxies to show nuclear activity. We make use
of the black hole mass -- bulge luminosity (M$_{BH}$ -L$_{bulge}$) relation, derived for
nearby inactive ellipticals, to estimate the central black hole  mass in our sample of
radio loud active galaxies. In spite of a $\sim$ 2 order of magnitude  difference of
intrinsic nuclear luminosity BLL and RLQ  have BH of similar mass ($<M_{BH} /M_{\odot}>
_{BLL}$= 5.6$\times$ 10$^{8}$, $<M_{BH}/M_{\odot}>_{RLQ}$= 1.0$\times$ 10$^{9}$). This
implies that the two types of objects are radiating at very different rates with respect
to their Eddington luminosity.
\end{abstract}

\begin{keywords}
RLQ and BL Lacs - active galaxies
\end{keywords}

\section{Introduction}
There is a general consensus  about the existence  of supermassive black holes (SBH) at
the center of  normal galaxies as well as in the nuclei of active galaxies and quasars
(see e.g. the recent review of Ferrarese 2002). A large body of data, in particular based
on high resolution HST observations, is now available to strongly support the presence of
such massive BH.

SBHs  play an important role in the formation and evolution of massive galaxies
and are also a key component for the development of the nuclear activity. In spite of
this apparently ubiquitous presence of SBH in galaxies our understanding on  how the
galaxies and their central BHs are linked in the process of formation of the structures
remains unclear but several attempts of explanation have been proposed (e.g. Silk \&
Rees 1998; Haehnelt \& Kauffmann 2000; Adams et al 2001; Burkert \& Silk 2001; Merritt
\& Ferrarese 2001a; Balberg \& Shapiro 2002).

From the observational point of view it was shown that the BH mass is correlated with
the properties of the bulge component of the host galaxy, which is translated into the
relationships between M$_{BH}$ and the bulge luminosity (Kormendy \& Richstone 1995;
Magorrian et al. 1998; Richstone et al. 1998; Kormendy \& Gebhardt 2001) and between
M$_{BH}$ and the velocity dispersion $\sigma$ of the host galaxy (Ferrarese \& Merritt
2000; Gebhardt et al. 2000;  Merritt \& Ferrarese 2001b). Any theory of SBH and galaxy
formation must therefore take into account and explain such observed empirical
relations (e.g. Silk \& Rees 1998, Haehnelt \& Kauffmann 2000; Ciotti and van Albada 2001).
On the other hand, although these relations have a significant scatter ( $\sim$0.4 in
Log M$_{BH}$), they offer a new tool for evaluation of BH masses in various types of
AGN if a reliable measurement of the host galaxy luminosity or of $\sigma$  is done.
While for AGN with strong emission lines (as QSO and Seyfert galaxies) the standard
methods (e.g. reverberation mapping) under virial assumptions of the emitting regions
can be used to derive M$_{BH}$ ( see e.g. Wandel et al. 1999 and Kaspi et al. 2000), the
above relations may be  the only way to estimate M$_{BH}$ for active galaxies that lack
of emission lines or that are too far away 
(as BL Lac objects and many nearby radiogalaxies)
to resolve the region of influence of the BH with present-day instrumentation. The two
different approaches lead to consistent estimates of BH masses within the assumed
uncertainties of the two methods (McLure \& Dunlop 2002).

Given the difficulty to obtain $\sigma$ from spectroscopy of the galaxies hosting active
nuclei, only for few AGN it was possible to use $\sigma$ to evaluate M$_{BH}$
(Ferrarese et al. 2001, Barth et al. 2002,2003; Falomo et al. 2002,2003). On the contrary
the galaxy luminosity is much easier to measure for active galaxies and can therefore be
used to determine M$_{BH}$ for larger data set.

In this paper we use the M$_{BH}$ -- L$_{bulge}$ relation to investigate and compare
the BH mass distribution of a sample of low and high luminosity radio-loud AGNs (BL
Lacs and RLQs respectively). Both classes  are found to reside in massive giant
ellipticals (Urry et al. 2000; Dunlop et al. 2001), which makes them rather homogeneous
for such kind of analysis. To ensure uniformity of the results we have considered only
objects at z$<$0.5 and that have been imaged by HST. This allows us also to better
constrain  their host properties. In section 2 we describe our samples of BLLs and RLQs
and compare their host galaxy luminosity functions. In section 3 we discuss the M$_{BH}
$-L$_{bulge}$ relation and derive the central black hole mass for each object. Finally
in section 4 we discuss our findings comparing with recent results on radiogalaxies. 
 In our analysis H$_{0}$=50 Km s$^{-1}$ Mpc$^{-1}$ and $\Omega_{0}$=0 were
used.

\section[]{Luminosity  of the host  galaxies}

We have collected host galaxy data for BL Lacs and RLQs at z$<$0.5 imaged 
by HST with the WFPC2 and have constructed a  homogeneous
dataset of the  host galaxies luminosities. This yields a sample 57 BL Lacs 
and 18 RLQs that represent, respectively, low  and high luminosity 
radio loud active galaxies.

Since most of the observations were obtained in the F702W and F675W filters we
converted all HST magnitudes into R Cousins band (Holtzman et al. 1995). In the few
cases where  the filters F555W,  F606W were used we applied a color correction V-R =
0.61 for the elliptical host galaxies (Fukugita et al. 1995). Absolute magnitudes have
been k-corrected following Poggianti (1997) prescriptions and corrected for galactic
reddening using the Bell Lab Survey of neutral hydrogen N$_{H}$ (Stark et al. 1992) with 
the conversion logN$_{H}$/E(B-V)=21.83 cm$^{-2}$ mag$^{-1}$ (Shull \& van
Steenberg 1985) assuming a total-to-selective extinction A$_{R}$=2.3E(B-V) (Cardelli,
Clayton \& Mathis 1989). Since the objects are distributed over a significant redshift
interval  we have also applied a correction to set the host galaxy luminosity to
present epoch assuming a passive stellar evolution for massive ellipticals (Bressan et
al. 1994). This correction ($\Delta$m $\sim$ -0.2 ) allows us to properly use the
M$_{BH}$-L$_{bulge}$ relations which refers to local galaxies. In the following M$_{R}$
represents the host galaxy absolute magnitude including all correction terms specified
above.
 
\subsection{The BL Lac Objects sample}

The HST snapshot image survey of BL Lacs (Urry et al. 2000, Scarpa et al. 2000) has
provided a homogeneous set of 110 short exposure high resolution images through the
F702W filter. From this we have extracted all resolved objects at z$<$0.5, yielding 57
sources with redshift between 0.027 and 0.495 ($<z>$=0.20$\pm$0.11). The  host galaxy
morphology of these objects is always well described by an elliptical model (Scarpa et
al. 2000). The absolute  M$_{R}$ magnitude for each object is reported in Table 1. The
host galaxy average luminosity is  $<M_{R}>$=-23.49$\pm$0.5, roughly one magnitude
brighter than the characteristic  galaxy magnitude M$_R^{*}$= -22.75 (Metcalfe et al.
1998). According to the shape of their spectral energy distribution (SED), BL Lacs are
broadly distinguished into two types (see Padovani \& Giommi 1995) : those the SED of
which peaks at near-infrared/optical and the $\gamma$-ray MeV regions (low frequency
peaked BL Lacs or LBL), and those that have SED peaking in the UV/X-ray and the
$\gamma$-ray TeV energies (called high frequency peaked BL Lacs or HBL). As shown by
Urry et al. 2000 the host galaxy properties of HBL  and LBL Lacs are indistinguishable,
and therefore the two subclasses will not be separated for this analysis.

\begin{table}
\caption{Host Galaxies Properties of BL Lac Objects: (a) name of the source, 
(b) redshift, (c) absolute R host galaxy magnitude, 
(d) black hole mass in units of M$_{\odot}$ in logarithmic scale. 
For reference see Urry et al. 2000, Scarpa et al. 2000.} 
\centering
\begin{tabular}{lccc}
Objects & z
  & M$_{R}$ & M$_{BH}$ \\
(a) & (b) & (c)  & (d) \\ \hline
0122+090  &  0.339     &  -23.45&   8.73       \\ 
0145+138  &  0.124     &  -22.61&   8.31        \\ 
0158+001  &  0.229     &  -22.84&   8.42        \\ 
0229+200  &  0.139     &  -24.47&   9.24      \\ 
0257+342  &  0.247     &  -23.83&   8.92      \\ 
0317+183  &  0.190     &  -23.53&   8.77       \\ 
0331--362 &  0.308     &  -24.06&   9.03      \\ 
0347--121 &  0.188     &  -23.02&   8.51      \\ 
0350--371 &  0.165     &  -23.19&   8.60    	\\ 
0414+009  &  0.287     &  -24.59&   9.30    	\\ 
0502+675 &   0.314     &  -23.60&   8.80      \\ 
0506--039&   0.304     &  -23.53&   8.77      \\ 
0521--365&   0.055     &  -23.24&   8.62      \\  
0525+713 &   0.249     &  -24.26&   9.13       \\ 
0548--322&   0.069     &  -23.63&   8.82   	\\  
0607+710 &   0.267     &  -24.10&   9.05       \\ 
0706+591 &   0.125     &  -23.90&   8.95     \\ 
0737+744 &   0.315     &  -24.04&   9.02        \\ 
0806+524 &   0.138     &  -23.39&   8.70    \\ 
0829+046 &   0.180     &  -23.64&   8.82          \\ 
0927+500 &   0.188     &  -22.96&   8.48       \\ 
0958+210 &   0.344     &  -23.32&   8.66        \\ 
1011+496 &   0.200     &  -23.41&   8.71    \\ 
1028+511  &  0.361     &  -23.75&   8.88     \\ 
1104+384   & 0.031     &  -23.15&   8.58     \\
1133+161  &  0.460     &  -23.33&   8.67       \\ 
1136+704   & 0.045     &  -22.84&   8.42         \\
1212+078  &  0.136     &  -23.79&   8.90     \\ 
1215+303   & 0.130     &  -23.66&   8.83         \\ 
1218+304   & 0.182     &  -23.38&   8.69        \\ 
1221+245   & 0.218     &  -22.29&   8.15     \\ 
1229+643   & 0.164     &  -23.91&   8.96     \\ 
1248--296  & 0.370     &  -23.81&   8.91      \\ 
1255+244  &  0.141     &  -23.17&   8.59       \\ 
1407+595  &  0.495     &  -24.32&   9.16        \\ 
1418+546  & 0.152      &  -23.93&   8.97    \\ 
1426+428  &  0.129     &  -23.54&   8.77        \\ 
1440+122  &     0.162  &  -23.52&   8.76       \\ 
1458+224  &     0.235  &  -23.48&   8.74     \\ 
1514--241  &    0.049  &  -23.48&   8.74       \\
1534+014  &     0.312  &  -23.93&   8.97    \\ 
1704+604   &    0.280  &  -22.95&   8.48      \\ 
1728+502  &     0.055  &  -22.32&   8.16       \\  
1749+096   &   0.320   &  -23.32&   8.66       \\ 
1757+703  &     0.407  &  -23.26&   8.63     \\ 
1807+698  &     0.051  &  -23.89&   8.95    \\  
1853+671  &     0.212  &  -22.96&   8.48        \\ 
1959+650  &     0.048  &  -23.06&   8.53        \\
2005--489 &     0.071  &  -23.78&   8.89      \\  
2007+777  &     0.342  &  -23.59&   8.80     \\ 
2143+070  &     0.237  &  -23.46&   8.73       \\ 
2200+420  &     0.069  &  -23.53&   8.77     \\  
2201+044  &     0.027  &  -22.51&   8.26    \\
2254+074  &     0.190  &  -24.23&   9.12    \\ 
2326+174  &     0.213  &  -23.47&   8.74    \\ 
2344+514  &     0.044  &  -24.13&   9.07    \\
2356--309  &    0.165  &  -23.06&   8.53        \\ 
\hline \\
\end{tabular}
\end{table}

\subsection{The RLQs sample}

Since there is not a homogeneous and  large set of HST observations for RLQs,  we
 have constructed a sample of 18 RLQ from the merging of three different subsets (BK:
 Bahcall et al. 1997 and  Kirhakos et al. 1999; Bo: Boyce et al. 1998; D: Dunlop et
 al. 2003). Bahcall et al. 1997 and  Kirhakos et al. 1999 studied 8 RLQ in the F606W
 and F555W filters and in the redshift range 0.158$<z<$0.367; Boyce et al. 1998
 reported the analysis for 5 sources with 0.223$<z<$0.389 in the F702W filter. The
 largest subsample was investigated  by  Dunlop et al. (2001), who report host
 galaxy properties for 10 radio-loud quasars with 0.1$<z<$0.25 observed in the F675W
 filter. As in the case of BLL  an elliptical model  is always a good representation
 for the host galaxies. 
 The average properties of the three subsamples are reported in
 Table 2. Our evaluations of M$_{R}$ are consistent with absolute values reported by
 the quoted authors when galactic extinction, filter correction and evolution
 correction are taken into account. 
 Since these subsets have statistically indistinguishable
 host luminosity distributions we have merged these subsamples to construct a
 representative sample of RLQ (see also Treves et al. 2002), taking average values
 for objects observed twice. The combined dataset consists therefore of 18 objects
 with redshift in the range 0.158$<z<$0.389, $<z>$=0.26$\pm$0.07 and
 $<M_{R}>=-24.04\pm0.4$. In Table 3 we give for each source the redshift z, the
 assumed galactic extinction in the R-band A$_{R}$, the host galaxy apparent
 magnitude R and the absolute magnitude M$_{R}$. In figure 1 we compare the redshift 
 distributions for the RLQ and BLL samples.

\begin{figure}
\psfig{figure=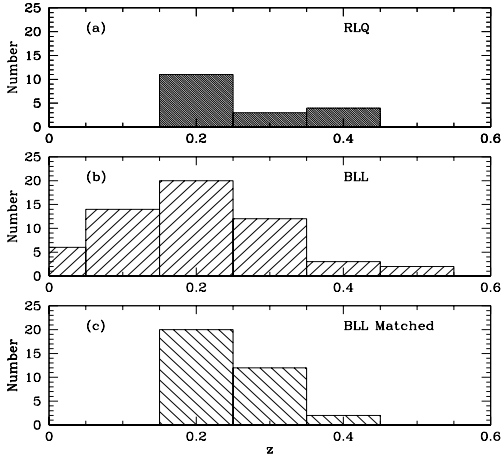,height=6.5cm,width=7cm}
\caption[]{Redshift Distribution for RLQ (a), BLL 
(b) and BLL matched samples (see text) (c).}
\label{eps1}
\end{figure}

\begin{table}
\caption{Average properties of three subsample of RLQs: 
(a) The sample (BK=Bahcall et al. 1997 and  Kirhakos et al. 1999; 
Bo=Boyce et al. 1998;  D: Dunlop et al. 2001); (b) Number of objects; 
(c) average redshift; (d) average absolute host galaxy magnitude.
}
\begin{tabular}{llll}
Sample  & N     & $<z>$ &   $<M_{R}>$    \\
(a)	& (b)	&(c)	&   (d)	      \\ \hline
BK	& 8	&0.26	& -23.92$\pm$0.61 \\
Bo	& 5	&0.30	& -24.23$\pm$0.27\\
D	& 10	&0.22	& -24.02$\pm$0.29\\
\hline \\
\end{tabular}
\end{table}

\begin{table}
\caption{Host Galaxies Properties of RLQs: (a) name of the source, 
(b) redshift z, (c) adopted extinction in the R band A$_{R}$, 
(d) apparent host galaxy R magnitude, (e) absolute R host galaxy magnitude, 
(f) black hole mass in units of M$_{\odot}$ in logarithmic scale and 
(g) reference, where BK=Bahcall et al. 1997 and Kirhakos et al. 1999, 
Bo= Boyce et al. 1998, D=Dunlop et al. 2003.} 
\begin{tabular}{lllllll}
Objects & z & A$_{R}$ 
   & R   
& M$_{R}$  & M$_{BH}$ & Ref. \\
(a) & (b)  &   (c)   &  (d) & (e)  &  (f) &   (g) \\ \hline
3C48       &   0.367&  0.26   & 17.2    &    -24.88  &   9.44 	& BK\\
PHL 1093       &   0.258&  0.14   & 17.2  &    -24.22  & 9.11 	& D/Bo \\
PKS 0202-76     &   0.389&  0.94   & 19.5 &   -23.83  &  8.92 	&Bo \\	
0312-77     &   0.223&  0.76   & 17.1    &    -24.58  &  9.29 	& Bo\\
0736+017     &   0.191&  0.46   & 16.9    &   -24.08  &  9.04  	& D\\
1004+130     &   0.240&  0.16   & 16.9    &   -24.26  &  9.13  	& BK/D \\
1020-103    &   0.197&  0.24   & 17.2     &    -23.64  & 8.82 	& D\\
3C273      &   0.158&  0.07   & 15.8   &      -24.20  &  9.10 	 & BK\\
1217+023    &   0.240&  0.08   & 17.3    &    -23.88  &  8.94 	 &D \\
1302-102    &   0.286&  0.09   & 17.8     &   -23.64  &  8.82  	&BK/Bo \\
B21425+267    &   0.366&  0.09   & 17.8     &  -24.05  & 9.03  	& BK\\
3C323.1     &   0.264&  0.19   & 18.1    &    -23.18  &  8.59   	& BK\\
3C351      &   0.371&  0.09   & 18.0    &     -24.36  &  9.18 	& Bo\\
2135-147    &   0.200&  0.34   & 17.4     &    -23.58  & 8.79  	& BK/D\\
OX169      &   0.213&  0.36   & 17.2     &     -23.95  & 8.98 		& D\\
2247+140     &   0.237&  0.24   & 17.2    &    -24.10  & 9.05     	&D \\
2349-014    &   0.173&  0.16   & 16.0     &    -24.42  & 9.21 	& BK/D\\
2355-082    &   0.210&  0.14   & 17.1    &     -23.80  & 8.90  	 & D\\    
\hline \\
\end{tabular}
\end{table}

\subsection{Comparison of the host luminosity of BLL and RLQ }

Both BLL and RLQ in our sample have been mostly discovered as counterparts 
of radio and/or X-ray sources.
Therefore the  objects considered here were selected on the basis of the nuclear properties, and
because there is not a significant correlation between the nuclear and host galaxy
luminosity (Urry et al. 2000; Percival et al. 2001; Dunlop et al 2003 ), we can consider
the distribution of the host galaxy luminosity is unbiased by selection
effects. Moreover the  homogeneous treatment of the data attests a
reliable comparison of host luminosity between the two classes (BLLs and RLQs). The
main limitation of this comparison remains however the exiguity of the RLQ sample.

We find that the average absolute magnitude of RLQ is about 0.5 magnitude brighter than
that of BLL. The difference is illustrated in figure 2 where we compare the cumulative
absolute magnitude distributions of the hosts for the two samples (a KS test indicates
that they are  statistically different at the $>$ 99\% level). Since the two samples span
slightly different redshift range we  checked that this does not affect our result. If we
consider a subsample of BLL with redshift distribution matched with that of RLQ (see
figure 1 (c)) we find $<M_{R}>_{BLL}$ (matched)= -23.54$\pm0.47$ thus confirming our
finding. Given the homogeneity of data analysis and the  procedure for the  selection of
the objects we believe that this difference is not biased. We note, however, that a
larger number of objects (in particular of RLQ) is required to confirm this result on a
firm statistical basis.

To further compare the luminosity distributions of the host galaxies  we constructed the
host galaxy luminosity function (HGLF) for the two subsets of objects. To set the
normalization of the HGLFs we simply assume the space density of both class of objects as
derived from studies of complete samples. For BL Lacs we use the value $\Phi_{0}$=
10$^{-5}$ Mpc$^{-3}$mag$^{-1}$ of the FR I radiogalaxies luminosity function at M$_{R}
$=-22.8 given by Padovani and Urry (1991) under the assumption that FR I radiogalaxies
are the parent population of  BL Lacs (e.g. Urry and Padovani 1995). For RLQ, we took the
value of the LF of close-by radio quiet QSO (Koehler et al 1997, Grazian et al 2000) at
M$_{B}$= -25.1, which corresponds to the average value of the nuclear magnitude for our
sources, and scaled it by a factor 10 to account for the ratio between RQQ and RLQ (e.g.
Moderski et al 1998). This yields $\Phi_{0}$=2.3$\times$10$^{-9}$ Mpc$^{-3}$ mag$^{-1}$.

In figure 3 we show the HGLF of BLL and RLQ compared with that of inactive ellipticals
(Metcalfe et al. 1998). To quantify the differences in shape of the HGLF we fitted the
luminosity distributions  of the host galaxies with a modified Schechter function
$\Phi$=K $\times$ $\Phi_{S}$ $\times$ (L/L$^{*}$)$^{\beta}$, where $\Phi_{S}$ is the
Schechter function for elliptical galaxies (Metcalfe et al. 1998): $\Phi_{S}$=$\Phi^{*}
$ $\times$ (L/L$^{*}$)$^{\alpha}$ $\times$ exp(-L/L$^{*}$), assuming $\Phi^{*}=
8.5\times10^{-2}$ Mpc$^{-3}$, $\alpha$=-1.2 and L$^{*}$=2.25 $\times$ 10$^{44}$ erg
s$^{-1}$ (Metcalfe et al. 1998). The best fit to HGLF was estimated minimizing
$\chi^{2}$ for the function $\Phi$. We find $\beta = $2.7 $\pm$0.2 for BLLs,
$\beta = $3.6$\pm$0.3 for RLQ. The shapes of the two HGLF  are somewhat 
different, but only at the 2-$\sigma$ level. 

This suggests that a given elliptical has a probability of having a radio-loud active
nucleus depending on the galaxy luminosity.  Moreover one can argue  that the steepness
of this behavior depends on the intrinsic luminosity of the nucleus as hinted by the
different value of $\beta$ for BLL and RLQ.

It turns out therefore that both types of radio loud active galaxies exhibit a remarkable
different distributions with respect to normal ellipticals, with clear preference for
more luminous (and massive) galaxies to show nuclear activity. This behavior disagrees
with that found by Wisotzki et al. (2001) for the host galaxies of radio-quiet QSO. The
shape of HGLF of the latter objects in fact appears to be consistent with that of
ordinary inactive early type galaxies.

\begin{figure}
\psfig{figure=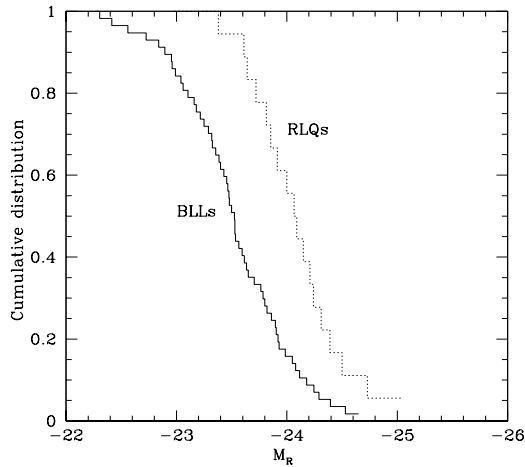,height=6.5cm,width=7cm}
\caption[]{Comparison of the cumulative host galaxy absolute magnitude (R band) 
distributions of RLQ (dotted line) and  BL Lacs (solid line).}
\label{eps1}
\end{figure}

\begin{figure}
\psfig{figure=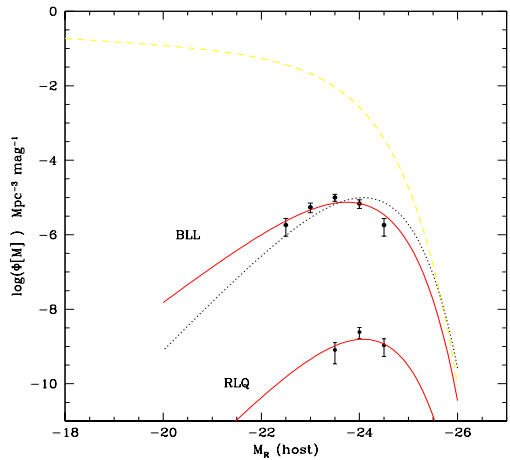,height=6.5cm,width=7cm}
\caption[]{The HGLF of RLQs and BL Lacs ({\it filled points}) compared with 
the fit ({\it solid lines} ) with a modified Schechter function (see text).
A slightly different value of $\beta$ is found for BLL ($\beta$ =2.7) and RLQ ($\beta$ =3.6). 
The dotted line is the fit to RLQ data normalized to the BLL data.
The {\it dashed curve} represents the luminosity function of elliptical galaxies  of 
Metcalfe et al. (1998). }
\label{eps2}
\end{figure}

\section{Mass of the Central Black Hole}

Basing on dynamical studies of nearby early type  galaxies it was shown that there  is a
linear relation between the luminosity of the spheroidal component of a galaxy
(L$_{bulge}$) and the mass M$_{BH}$ of the central black hole (e.g. Kormendy \& Gebhardt
2001 and references therein). This correlation has a  scatter  of $\sim$ 0.4 in
Log(M$_{BH}$) that can be ascribed  mainly to the  errors of measurements of M$_{BH}$ and
to the uncertainties to disentangle the bulge from the disc component of the galaxies.
Nevertheless it can be used to estimate M$_{BH}$ for our objects provided that 
consistent and reliable host galaxy luminosities are used.

To avoid systematic effects it is important that the adopted absolute magnitude
of the galaxy be homogeneous (in terms of spectral band, adopted cosmology, extinction
correction, filter, etc) with that used to derive the M$_{BH}$-L$_{bulge}$ relation. To
satisfy this requirement we used the relationship between M$_{BH}$ and M$_R$ derived by
Bettoni et al. (2003) for 20 inactive ellipticals assuming the same calibrations of
M$_{R}$ we have adopted here.

From the analysis of the inactive sample of 
ellipticals Bettoni et al. give 

\begin{equation}
log(M_{BH}/M_{\odot})=-0.50\times M_{R}-3.00
\end{equation}

that is used to derive M$_{BH}$ from absolute (total) magnitude M$_R$ of 
ellipticals (H$_{0}$=50 Km s$^{-1}$ Mpc$^{-1}$). 
This relation has  rms scatter of 0.38 in log(M$_{BH}$) and it 
is similar to that derived by McLure \& Dunlop (2002).

Since the host galaxies in the samples considered here are all bona-fide ellipticals we
used  relation (1)  to derive M$_{BH}$ for the two samples of radio loud AGN (BL Lacs and
RLQs) and report the value for each object in table 1 and 2, respectively. The
uncertainty  on the estimated black hole mass is dominated  by the scatter of  relation
(1) while the uncertainties on the host galaxy magnitude  are usually smaller.

The distributions of M$_{BH}$ for the two samples are shown in Figure 4. The two classes
exhibit an average difference by a factor $\sim$ 2 in M$_{BH}$ as a consequence of the
different average host luminosity. We find that the average values of M$_{BH}$ are
$<log(M_{BH}/M_{\odot})>=8.75\pm$0.25 and $<log(M_{BH}/M_{\odot})>=9.02\pm$0.20
respectively for BLLs and RLQs.

\begin{figure}
\psfig{figure=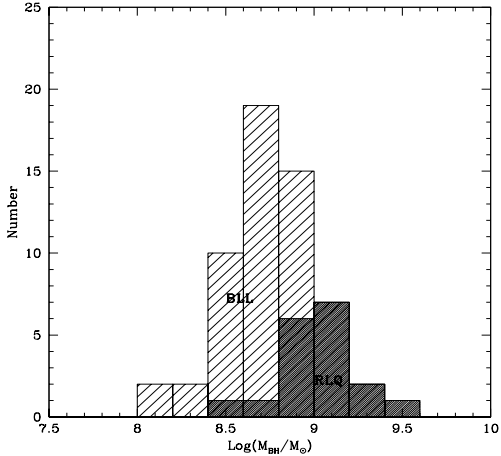,height=6.5cm,width=7cm}
\caption[]{Comparison of the Black Hole Mass (derived from the host galaxy luminosity using the 
relation between M$_{BH}$ and bulge luminosity of nearby ellipticals) distributions of
RLQs and BLLs}
\label{eps3}
\end{figure}

A complementary method to derive BH masses is based on the relation between the BH masses
and $\sigma$ of the host galaxy (Gebhardt et al. 2000 and Ferrarese et al. 2000). This
has been applied to a small number of nearby BL Lac objects (Falomo et al. 2002,2003;
Barth et al. 2002;2003). As shown by Falomo et al 2003 there is a good agreement in the
results obtained with the two techniques. Thus no estimates of BH mass of RLQ
are available from $\sigma$ because of the lack of measurements.

Wu, Liu and Zhang (2002) and Woo and Urry (2002) have derived estimates of the BH mass of
BL Lacs using the M$_{BH}$ -- $\sigma$ relation where the latter quantity was inferred
from measurements of  the effective surface brightness and the effective radius of the
host galaxy and  by assuming these are linked to $\sigma$ through the Fundamental Plane
relationship. Although in principle this method could work we believe it is 
less accurate than the direct use of the M$_{BH}$ -- M(bulge) relation. In fact  in
addition to the uncertainty in the measured quantities ($\mu_e$ and R$_e$), which are
much larger than the total magnitude of the galaxy, one has to take into account the
uncertainty due to the intrinsic scatter of the FP relation and that of the M$_{BH}$ --
$\sigma$ relation. Comparison of our BH masses (see Table 1) with those derived by Woo
and Urry or Wu et al 2002 indicate remarkable differences of M$_{BH}$ for many objects
mainly because of the wrong evaluation of the velocity dispersion. In some cases (e.g.
1807+698, 1104+384) the poor estimate of the velocity dispersion via the FP method has been
clearly confirmed by direct measures of $\sigma$ (Barth et al; Falomo et al).

We also note that in  the Woo \& Urry (2002) estimates additional  errors derive from a
mistreating of the FP parameters because instead of the average surface brightness
($<\mu>_{e}$), the isophotal surface brightness was used.

\section {Discussion}

The analysis of HST images for low redshift BL Lacs and radio loud quasars has shown that
for both types of active nuclei the host galaxies are very luminous ellipticals. On
average they are $\sim$1-2 mag more luminous than the typical galaxy luminosity (M$^{*}
_R\sim$ --22.75; Metcalfe et al. 1998). After  homogeneous treatment of the data we
also found that host galaxies of RLQ are systematically  more luminous by $\sim$0.5
magnitudes than BL Lac hosts. Although this result does not seem to depend on the
selection of the objects, a larger  sample of RLQ is needed to reach a firm conclusion.
We have shown that the distribution of the host galaxy luminosity exhibits a marked drop
towards less luminous galaxies and that it is somewhat different for the two classes (BLL
and RLQ). This indicates that such kind of nuclear activity occurs preferentially (or
lasts longer) in massive galaxies.  The different distributions of host luminosity for
BLL and RLQ may simply reflect the very large range of  intrinsic nuclear luminosity
(about two orders of magnitude, see below). High power nuclear activity like that
observed in  RLQ  can occur only in the most luminous and massive galaxies and it is
therefore a rare event. On the other hand, low power nuclear activity as that observed in
BL Lacs (or in radio galaxies believed to be identical objects non affected by beaming
effects) can be present also in galaxies with intermediate luminosities.

On the assumption that the galaxy luminosity is correlated with the central BH mass, the
host galaxy luminosity  can be translated into central  BH masses. It turns out thus
that, within a factor of two, BLL and RLQ have similar  BH masses but their total
intrinsic nuclear luminosities are remarkably different. In addition to the higher
observed nuclear/host ratio of RLQ with respect to BLL we have to take into account the
fact that, while we consider RLQ basically unbeamed, for BLL a substantial beaming factor
is present ($\delta\sim$15 see Ghisellini et al. 1998, Capetti \& Celotti 1999). The
intrinsic nuclear luminosities therefore differ by about a factor 100. This implies a
dramatic difference of the Eddington ratio $\xi_{E}=L/L_{E}$ where L$_{E}$=
1.25$\times$10$^{38}\times$(M$_{BH} $/M$_{\odot}$) erg s$^{-1}$ (see also O'Dowd et al
2001, 2002; Treves et al 2002). Basing on the estimated total QSO luminosity of L$\sim
3\times$ 10$^{12} L_{\odot}$ (e.g. Elvis et al. 1994) and assuming  BH masses of 1-5 x
10$^{9}$ M$_{\odot}$, we find that RLQ may be emitting at rates of 10\% or higher than
their Eddington power, while BLL are always emitting at regimes that are much lower than
L$_{E}$.

\begin{figure}
\psfig{figure=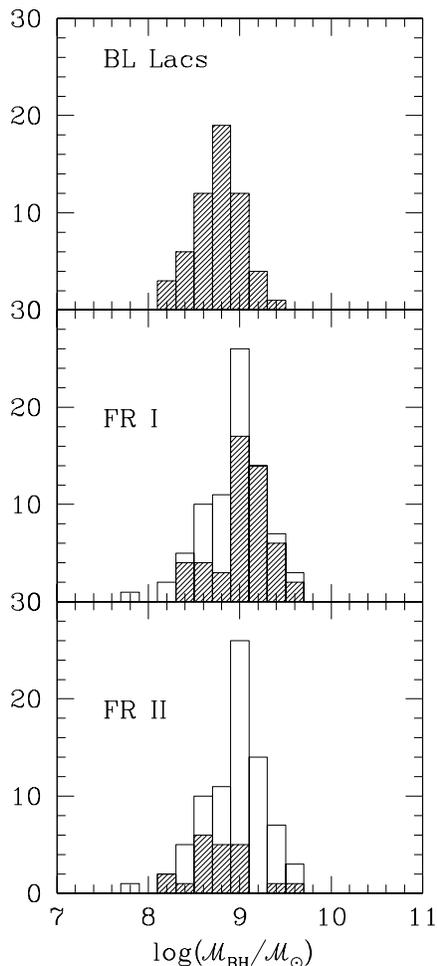,width=7cm}
\caption[]{Black hole mass distribution of our sample of BL Lacs ({\it top panel}) 
compared with that of FR I ({\it middle panel}) and FR II 
({\it bottom panel}) radiogalaxies studied by Bettoni et al (2003). 
The open histograms in the middle and 
bottom panels refer to the distribution of the whole sample of radiogalaxies (see text).}
\label{eps5}
\end{figure}

According to the unification schemes of radio loud AGN (e.g. Urry and Padovani 1995) BL
Lacs are radiogalaxies the jet of which is closely oriented toward the observer. Based on
arguments of number  density, luminosity functions and unbeamed  properties (as the
extended radio luminosity or the host galaxies) the parent population of BL Lacs is
likely formed by FR I radio galaxies with some contamination by FR II sources (Padovani
and Urry 1990; Wurtz et al 1996; Falomo and Kotilainen 1999; Cassaro et al 1999; Urry et
al 2000). Under this hypothesis the BH mass of BL Lacs ($<log(M_{BH}/M_{\odot})>=
8.75\pm$0.25 ) and of the parent (unbeamed) objects must be identical. In figure 5 we
compare the distribution of the BH mass for our sample of BL Lacs with that of low
redshift radiogalaxies from the sample of Govoni et al 2000 (see also Bettoni et al
2003). The two distributions are rather similar although the most massive BH in luminous
FR I radiogalaxies do not appear to have counterparts in the known BL Lacs. The average
values of M$_{BH}$ are $<log(M_{BH}/M_{\odot})>=9.04\pm$0.30 and $<log(M_{BH}
/M_{\odot})>=8.78\pm$0.35 for, respectively, FR I and FR II radio galaxies.

Finally we wish to note that in addition to the mass the other parameter which
characterizes a black hole and that may play a relevant role in the observed
phenomenology is the BH  spin. It has been suggested that the spin energy is responsible
for the jet emission L$_j$ and therefore for the development of the radio-emission (e.g
Blandford 2000, Dunlop et al 2003). The spin  is clearly not directly measurable but it
could be deduced from an estimate of L$_j$ and the BH mass using  the Blandford and
Znajek (1977) formula. While L$_j$ could be  obtained from the spectral energy
distribution (e.g. Tavecchio et al 2000, 2002), the BH mass may come through the
procedures described in this work.

\section*{Acknowledgments}
This work has received partial 
support under contracts COFIN 2001/028773, ASI-IR-115 and ASI-IR-35.

\bsp

\label{lastpage}

\end{document}